\documentstyle[aps,twocolumn,prl,epsf]{revtex}

\begin{document}
\newcommand{\be}{\begin{equation}}
\newcommand{\ee}{\end{equation}}
\newcommand{\bea}{\begin{eqnarray}}
\newcommand{\eea}{\end{eqnarray}}
\newcommand{\hn}{H\'{e}non }
\newcommand{\pc}{Poincar\'{e} }
\newcommand{\pcss}{Poincar\'{e} surface of section }

\newcommand{\bfx}{{\bf x}}
\newcommand{\bfC}{{\mathbf{C}}}
\newcommand{\bfF}{{\mathbf{F}}}
\newcommand{\bfes}{{\mathbf{e}}_{s,i}}
\newcommand{\bffs}{{\mathbf{f}}_{s,i}}
\newcommand{\bfeu}{{\mathbf{e}}_{u,i}}
\newcommand{\eu}{{\mathbf{e}}_{u}}
\newcommand{\es}{{\mathbf{e}}_{s}}
\newcommand{\bffu}{{\mathbf{f}}_{u,i}}
\newcommand{\bfesi}{{\mathbf{e}}_{s,i}}
\newcommand{\bffsi}{{\mathbf{f}}_{s,i}}
\newcommand{\bfeui}{{\mathbf{e}}_{u,i}}
\newcommand{\bffui}{{\mathbf{f}}_{u,i}}
\newcommand{\bff}{{\mathbf{f}}}
\newcommand{\bfg}{{\mathbf{g}}}
\newcommand{\bfU}{{\mathbf{U}}}
\newcommand{\bfV}{{\mathbf{V}}}
\newcommand{\bfW}{{\mathbf{W}}}
\newcommand{\bfX}{{\mathbf{X}}}

\newcommand{\dpmax}{\delta p_{\rm max}}
\newcommand{\dpn}{\delta p_n}
\draft
\bibliographystyle{prsty}
\newcounter{romantime}
\title{Control and Tracking of Unstable Orbits in Hamiltonian Flows:\\
         The Diamagnetic Kepler Problem}
\author{Babak Pourbohloul and Louis J. Dub\'e\cite{e-mail}}
\address{D\'epartement de Physique, Universit\'e Laval, Cit\'e Universitaire, Qu\'ebec, Canada
G1K 7P4}
%\date{submitted to {\em Phys. Rev. Lett} 1 March 2000, revised 10 May 2000}
\maketitle

\begin{abstract}
We report on the application of chaos control to the irregular motion of an electron under the combined
influence of a Coulomb and a magnetic field, the so-called ``diamagnetic Kepler problem'' (DKP).
We show how to stabilize the classical chaotic orbits into regular motion and present a new method to follow
the unstable orbits as the energy is increased from the regular to the chaotic regime.
\end{abstract}


%\narrowtext

The dynamics of an hydrogen atom in a strong magnetic field is known as the ``diamagnetic Kepler problem'' (DKP). After more than two decades of research, the DKP is currently the most intensively studied
autonomous, classically chaotic, atomic physics system.  As first realized by Edmonds \cite{EDM73},
the DKP is important in such diverse areas as solid state physics, astrophysics, and
Rydberg atoms and occupies central stage in classical and quantum chaos research \cite{BLU97}.  
The classical flow covers a wide range of Hamiltonian dynamics reaching from bound, nearly integrable behavior to completely chaotic and unbound motion as the scaled energy is varied \cite{dkp}. This letter reports  the numerical implementation of a control scheme to stabilize this chaotic behavior. The methodology is sufficiently
general to apply to {\em all} Hamiltonian flows and , with slight modifications, is also applicable to ballistic
dynamics (e.g. billiards) and area-preserving mappings.

Ever since Ott, Grebogi and York (OGY) \cite{ott90} introduced the notion of chaos control a decade ago, 
the number of experimental verifications has continuously and rapidly expanded to include a wide
variety of applications ranging from magnetoelastic ribbons \cite{ditto2}, electronic circuits \cite{hunt} , 
lasers \cite{roy} , chemical reactions \cite{petrov2} to thermal convection loops \cite{singer}. 
The OGY algorithm was written explicitly for dissipative systems and most studies (experimental and
theoretical) have focused their attention on this class of systems. 
Non-trivial modifications are needed to adapt the OGY technique to conservative flows (or maps)
and Lai, Ding and Grebogi \cite{lai93} have provided the necessary changes in the context of area-preserving
mappings.

Our  system is a continuous, 2 degrees of freedom (4D phase space) Hamiltonian flow
representing the motion of an electron under the combined influence of a Coulomb and an external
magnetic field.  We use scaled semi-parabolic coordinates and write the resulting scaled (pseudo-) Hamiltonian as  (for angular momentum $L=0$) \cite{friedrich} 
\be
   \hat{h}_{DK} = \frac{1}{2}(p^2_{\nu} + p^2_{\mu}) -{\epsilon}\, (\nu^2 + \mu^2)
                  + \frac{1}{8} \nu^2 \mu^2 (\nu^2 + \mu^2) \equiv 2 \quad .
\label{hpseudo}
\ee 
The scaled energy $\epsilon$ is related to the physical energy $E$ by $\epsilon = \gamma_0^{-2/3}\ E$
where the parameter $\gamma_0 = B / B_0$ denotes the strength of the magnetic field relative
to the unit $B_0 \simeq 2.35\ 10^5\ T$.  The structure of the dynamics of the Hamiltonian (\ref{hpseudo}) depends solely on the value of the scaled energy $\epsilon$. For instance, it is known that for  $- 0.5 < \epsilon  < - 0.13$ the system exhibits bounded motion with mixed chaotic and regular motion and for $- 0.13 < \epsilon  < 0.0$ the last large stable island disappears and the dynamics is mostly chaotic \cite{tanner}.

The dimension reduction (from 4D to 2D) 
and discretization is performed by observing the dynamics on the Poincar\'e section defined
by $\mu= 0, \dot{\mu}> 0$. The energy shell is then mapped to an area bounded by the
condition $p_{\nu}^2 - 2 \epsilon\ \nu^2 = 4$ which represents an ellipse in the
$(\nu, p_\nu)$ plane. Figure (1) shows the collection of points $\{\nu_n, p_{\nu,n}\}$
obtained by numerical integration of the equations of motion for $\epsilon= -0.5$.
One notices, for this energy, that the {\em mixed} phase space is still mostly regular with a small stochastic
area filled by the successive piercings of {\em one} chaotic trajectory. We recall that a
periodic trajectory would show up on the Poincar\'e section as a finite number of crossings. 

In attempting to bring order to the DKP dynamics, we had to overcome a number of difficulties
not encountered in previous studies. 
We had in mind the {\em adaptive} control of chaotic DKP trajectories under changing conditions
(e.g. drift in the magnetic field) through the stabilization of preselected unstable periodic 
orbits (UPOs) (monitored on the Poincar\'e section) via {\em small} programmed perturbations 
to the scaled energy. The difficulties are five-fold.   
First, a typical trajectory spends a lot of time away
from the Poincar\'e section where the flow is discretized and because of the sensitivity 
of the chaotic dynamics we had to device 
an efficient variable step {\em symplectic} integrator thereby preserving the geometrical structure of the Hamiltonian \cite{POU00}.
Second, the locations of a number of UPOs (our target states) and the corresponding Jacobian matrices for each member of the periodic orbits must be obtained numerically. 
It was found  necessary to intervene at {\em every} crossing of the Poincar\'e section
to assure a robust stabilization.
Third, because of anomalous transport in Hamiltonian phase space \cite{MEI92}, the waiting time for a trajectory 
to enter the neighborhood of a predefined target UPO is unduly long and a {\em targeting} strategy is in
general necessary to steer the trajectory towards the appropriate region of phase space.  This problem
has been addressed in \cite{targeting} and is assumed  to be solved.  
Fourth, the eigenvalues
of  area-preserving Jacobians are often complex and the stable and unstable manifolds, at the core of the
OGY procedure, are no longer  along
the directions of their eigenvectors. A new method had to be implemented along the lines described
in \cite{lai93}. 
Fifth, a {\em tracking} algorithm for updating the control parameters (positions of UPOs, Jacobian matrices ...)  was designed to insure that control is kept while external conditions are changing. 
Part of our solutions to these problems are discussed below and the details will be reported elsewhere.
 
%
%  location of UPOs: modified recurrence method, STA
%

To locate the UPOs on the Poincar\'e section, we have used two methods: a modified recurrence method (MRM)
and an extension to conservative mappings of a stability transform
algorithm (STA) introduced in \cite{sta}. 
The MRM uses the basic idea proposed by Auerbach {\it et al.} \cite{auerbach} for extracting periodic orbits from a chaotic time series $\{ {\bf X}_i\} _{i=1,N} $. However, instead of scanning a very long $N \gg 1$ time series for pairs of points separated in time by $m$ iterates (for a period-m UPO) and within a preassigned spatial  distance $r$ from each other, $M$ points are launched on the Poincar\'e section and integrated
forward until the $m$-th return on the section where the pair of points are compared. In both methods,
the accepted pairs, a distance $r$ apart, are distributed among the possibly different
UPO candidates, the center of mass of each subset is calculated and the coordinates are stored.
For a given relative accuracy, $10^{-4}-10^{-6}$, we find $M \ll N$ by several orders of magnitude. In Hamiltonian
cases, $N$ must be at least $10^6$ in order to reach an acceptable precision.  In the MRM,     
one  can also easily take advantage of the symmetries on the Poincar\'e section for the flow. 
For example, figure (\ref{ps-05}) shows that the mapping is symmetric with respect to both $\nu $ and $p_\nu $ axes. Therefore one only needs to spread the initial conditions over one quarter of phase space for
a smaller set of initial conditions or a denser distribution. This optimization is essential if one considers
that between every two consecutive points on the Poincar\'e section, a segment of trajectory in phase space
must be integrated numerically. Furthermore, for the DKP, this segment grows rapidly as the scaled energy  approaches zero  where the trajectory spends an increasingly long time in the
branches (they grow as $|\epsilon|^{-1/2}$) of the pseudo-potential.

The STA amounts to the following.  Consider a discrete chaotic dynamical system $S$ defined
in a $d$-dimensional space by
\be
\bfX_{n+1} = \bfF(\bfX_n, p)\quad  \label{unstable}
\ee
where $p$ is some external parameter. The goal is to construct from equation (\ref{unstable}) other dynamical systems $S_k$ with the {\em same} periodic orbits (in number and positions), but whose
stability has changed, i.e unstable orbits become stable, and stable orbits remain stable. The new
systems are defined by the linear transformations (written for a period $m$ point, $\bfF^{(m)}(\bfX^*,p)= \bfX^*$)
\be
     \bfX_{n+1} = \bfX_n + \lambda\ \bfC_k\ [\bfF^{(m)}(\bfX_n, p) - \bfX_n]
\label{sta-trans}
\ee
where $0 < \lambda \ll 1$ is adjusted to improve convergence and $\bfC_k$ are non-singular constant
$d \times d$ matrices (see \cite{sta} for their explicit forms). 
Because of the stability of the periodic orbits of the constructed system $S_k$, every trajectory of $S_k$ converges after a finite number of iterations to a periodic $m$ point ${\bf X}^*$. Per construction, the
${\bf X}^*$ are also periodic points of the original system $S$.
We have tested this procedure for a number of area preserving mappings and flows and the global
convergence of the method is excellent. For flows, where the mapping in (\ref{sta-trans})
is replaced by a numerical integration for $m$ returns to the Poincar\'e section, the STA has
proved equally reliable. In this respect, our adaptive symplectic integrator \cite{POU00} has insured that the (long) trajectories away from the Poincar\'e section were kept sufficiently accurate.
The position of the fixed point
in figure (2) (bottom panel) has been reached after only 250 iterations 
starting from an arbitrary initial point: this is representative of the efficiency of the
STA. For the same relative accuracy ($10^{-6}$), the MRM needs approximately $10^4$ initial points.

%
%    control algorithm
%

In order to stabilize a chaotic trajectory around one of these UPOs,  we have implemented a numerical
version of the OGY method as modified in \cite{lai93} for area-preserving mappings. 
This method is believed to be dynamically optimal in that it explicitly uses the local geometry
of the underlying system.  Given  a dynamical system of the type (\ref{unstable}),  and a target UPO
of period $m$, $\{ \bfX(i,p_0) \}_{i=1,m}$ , at some nominal parameter value $p_0$, one characterizes
the local stable and unstable manifolds by the vectors  ${\bf e}_{s,i}$ and ${\bf e}_{u,i}$ 
respectively as well as their contravariant  counterparts ${\bf f}_{s,i}$ and ${\bf f}_{u,i}$ satisfying 
${\bf f}_{u,i}\cdot {\bf e}_{u,i} = {\bf f}_{s,i}\cdot {\bf e}_{s,i} = 1$ and ${\bf f}_{u,i}\cdot {\bf e}_{s,i} = 
{\bf f}_{s,i}\cdot {\bf e}_{u,i} = 0$.  The stabilizing perturbations $\delta p_n \equiv p_n -p_0$ are then obtained by firstly linearizing
the dynamics in a $\delta$-neighborhood of a member of the periodic orbit, say $\bfX(k,p_0)$ , and 
around $p_0$, namely
\be
    \bfX_{n+1} - \bfX(k+1,p_n) \sim 
{\bf U}_k\ \left[ \bfX_{n} - \bfX(k,p_n) \right]  
\label{eq-linear2d}
\ee
where the  $ d \times d$ Jacobian matrix $\bfU \equiv D_{\bfX} \ \bfF(\bfX,p)$   is
evaluated at $[\bfX = \bfX(k,p_0) , \  p = p_0]$ and it is understood 
that $||\bfX_{n} - \bfX(k,p_0)|| \leq \delta \ll 1$  and $|\delta p_n| \ll  |p_0|$. 

Secondly, the control criterion is imposed that $\bfX_{n+1} = \bfF(\bfX_n,p_n)$ should lie along
the {\em stable direction} at $\bfX(k+1,p_0)$, i.e.
$    \bff_{u,k+1} \cdot [\bfX_{n+1} - \bfX(k+1,p_0)]= 0 $
 which, together with the parametric variation of the periodic points, $\bfX(k, p_0 +\delta p) \sim
\bfX(k,p_0) + \bfg_k\ \delta p$ ,  leads to the following expression for the parameter perturbation at the 
$n$-th iteration:
\be
      \delta p_n = -\  \frac{  \bff_{u,k+1}  \cdot \left\{ {\bf U}_k\ \left[ \bfX_{n} - \bfX(k,p_0) \right]
                                \right\}  }
                      {  \bff_{u,k+1} \cdot (\bfg_{k+1} - \bfU_k\ \bfg_k)   } \quad .
\label{eq-dpng}
\ee

The individual Jacobian matrices $\bfU_i$ were obtained by least-square minimization \cite{jacobians} 
of the linear relationship between subsequent pairs of iterates on the Poincar\'e section for 
over 1000 points drawn from the close neighborhood of each $\bfX(i,p_0)$. 
For the dissipative cases, the eigenvectors of the Jacobian matrix
$\bfU_k^{(m)} = D_{\bfX} \bfF^{(m)}(\bfX_k,p_0) = \prod_{j=0}^{m-1} \bfU_{k+j}$ 
would determine the manifold directions, but as first realized in \cite{lai93}, the conservative cases lead
generically to complex eigenvalues and eigenvectors and a different strategy is required. The solution adopted
to generate the unstable (stable) direction $\eu$  ($\es$) is to rotate   an arbitrary unit vector ${\bf u}_0$
(${\bf s}_0$) according to $ \bfU^{(qm)}_k {\bf u}_0$  and $\bfU^{(-qm)}_k {\bf s}_0$
which converge respectively to ${\bf e}_{u,k}$ and ${\bf e}_{s,k}$ for $q \gg 1$. The  vectors
are periodically renormalized to prevent overflow and in general, 
$q \leq 10$ is sufficient for convergence. This procedure is very reliable and robust. 

%
%  Results
%%%%%%%%%%%%%%%%

Putting all the ingredients together, we have achieved the stabilization of the chaotic behavior of the 
diamagnetic-Kepler Hamiltonian (\ref{hpseudo}) for different values of $\epsilon$  and different UPOs. 
Figure (2)
%(\ref{dkp_p2_03, dkp_p3_02,dkp_p1_01})
presents a sample of results  for the control of period-2, period-3 and period-1 at $\epsilon$ = -0.3, -0.2 and -0.1 respectively.  The control mechanism was switched on when an iterate entered a neighborhood of size
 $\delta = 10^{-3} $ ( in the 3 panels, one has removed the initial transient) and the perturbations on
the scaled energy (alternatively the magnetic field) were never larger than 
$\delta \epsilon = 10^{-2}$.  After 3000 iterates, we release the control, $\delta \epsilon_n = 0$ , and the trajectory returns to its natural chaotic behavior. 
One also notes (left panels) that the dynamics is increasingly chaotic as $\epsilon $ approaches zero and by $\epsilon= -0.1$
no obvious trace of regularity is present.
We emphasize that  in performing the present control task no {\em a priori knowledge} of the governing map on the Poincar\'e section
is known and the complete procedure is purely numerical from beginning to end.

%
%  Adiabatic Localization 
% 
Under parameter variation, the location of the UPOs changes as well as their associated Jacobian 
matrices, and unless action is taken the control over the chaotic dynamics may be lost. The strategy
of detecting parameter drift and updating accordingly the control parameters is known as {\em tracking}
\cite{tracking}.  We have developed a new tracking algorithm which is simple and efficient. For concreteness,
we will concentrate the presentation on the updating localization of the UPOs. 

The method is based on a judicious construction of a family of {\em extrapolating rational functions}
\cite{press}. One starts from a number $s$ of UPO coordinates (the seeds),  e.g. 
$\{ \nu_i, \epsilon_i \}_{i=1,s}$,
from which a rational function, $R_0(s,\epsilon)$  is constructed.   $R_0$ is then evaluate at
$\epsilon_{s+1} = \epsilon_s + \Delta \epsilon$ with $\Delta \epsilon \ll 1$ to give an extrapolated
value $\nu_{s+1}$. A new set of seeds is formed $ \{ \nu_{i+1}, \epsilon_{i+1} \}_{i=1,s}$ , 
removing $\{ \nu_1, \epsilon_1 \}$ and keeping the number of seeds fixed , 
and a new extrapolating function $R_1(s,\epsilon)$ is calculated. The process is repeated until
a final value  $\epsilon_f$ is reached.  
The same procedure may also be applied to the pairs $\{ p_{\nu,i}, \epsilon_i \}_{i=1,s}$
to provide initial conditions $\{ \nu_i, p_{\nu,i} \}$ for the integration of the complete phase-space 
trajectory.
Two remarks are in order. First, this approach is
much more accurate at $\epsilon_f$ than the direct extrapolation $R_0(s,\epsilon_f)$, although
all the original seeds are replaced by extrapolated values after $s$ iterations. Second, the algorithm 
permits to cover a wide range of parameter variations beyond the initial seeds (3-5 in practice)
while keeping high accuracy.  
Since the individual extrapolation steps are chosen small, we call the method {\em adiabatic localization} (AL) or
more generally {\em adiabatic tracking} (AT).

We have applied  AT to the dissipative \hn 
map for the same range chosen in \cite{tracking}. The results are in complete agreement with 
those of \cite{tracking} and with the precise values of the position of UPOs obtained by STA. We have also applied this algorithm to the standard map,
the H\'enon-Heiles Hamiltonian and the DKP and more complete results will be reported at a later time. 
Figure (\ref{track}) shows the tracking of a period-1 orbit for the DKP. We started from five seeds for $\epsilon$: - 0.3, - 0.29, - 0.28, -0.27, -0.26 and set $\Delta \epsilon = 0.001$. The tracking process is extended to $\epsilon = -0.075$. The results from the STA confirms the accuracy of the adiabatic tracking. The right panel clearly indicates that the result from AT for $\epsilon = -0.075$ is indeed a period 1 orbit.
 
%
%  Conclusion
%
We have presented for the first time the application of a numerical implementation of the OGY
control to a realistic conservative flow, the DKP Hamiltonian. On our way to a complete solution,
we have i. developed  a new variable step symplectic integrator, ii. modified the standard recurrence
method and extended a powerful technique
(STA) to locate the positions of the unstable periodic orbits, iii. obtained accurate numerical Jacobian
matrices and the local manifolds, and iv. provided an efficient approach (AT)
for tracking the UPOs  under parameter changes. 
We are presently considering two important mesoscopic applications: 
nonlinear dynamics in electromagnetic traps \cite{PAU90}, and  chaotic ray dynamics in lasing
microcavities \cite{NOE97}.   
 It still remains an open question however if manipulations of 
an external parameter (e.g. a magnetic field) to induce stabilization of a 
classical unstable orbit can be extended to the semi-classical regime, for example,  for the control
of Rydberg wave packet dynamics. Research along these lines is actively being pursued.

%\bibliography{biblbank}
\renewcommand{\pre}[1] {{\it Phys.\ Rev.\  E} {\bf #1} }
\renewcommand{\prl}[1] {{\it Phys.\ Rev.\  Lett.} {\bf #1} }
\renewcommand{\pra}[1] {{\it Phys.\ Rev.\  A} {\bf #1} }
\renewcommand{\rmp}[1] {{\it Rev.\ Mod.\ Phys.} {\bf #1} }
\newcommand{\pla}[1] {{\it Phys.\ Lett.\  A} {\bf #1} }
\newcommand{\physicad}[1] {{\it Physica\ D} {\bf #1} }
\newcommand{\jpa}[1] {{\it J.\ Phys.\ A} {\bf #1} }
\newcommand{\jpb}[1] {{\it J.\ Phys.\ B} {\bf #1} }
\newcommand{\ibif}[1] {{\it Int.\ J.\ Bifurcation and Chaos} {\bf #1} }
\newcommand{\chaos}[1] {{\it Chaos} {\bf #1} }
%

%
%
%%%%  Figure Section with eps call-in sequences
%%%%%%%%%%%%%%%%%%%%%%%%%%%%%%%%%%%%%%%%%%%%%%%%%%%
\setcounter{figure}{0}
\begin{figure}[h]
\epsfxsize=.35\textwidth\leavevmode
\centerline{\epsffile{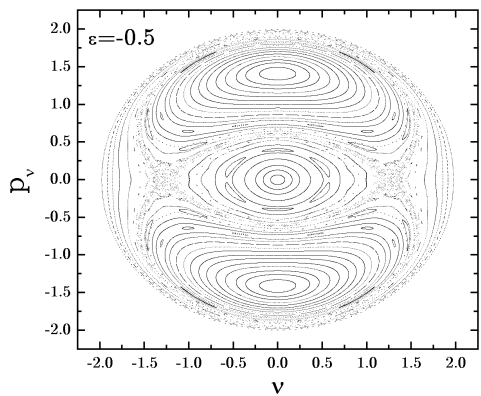}}
\caption{Poincar\'e section of the DKP for  $\epsilon$ = -0.5.}
\label{ps-05}
\end{figure}
\begin{figure}[h]
\epsfxsize=.45\textwidth\leavevmode
\centerline{\epsffile{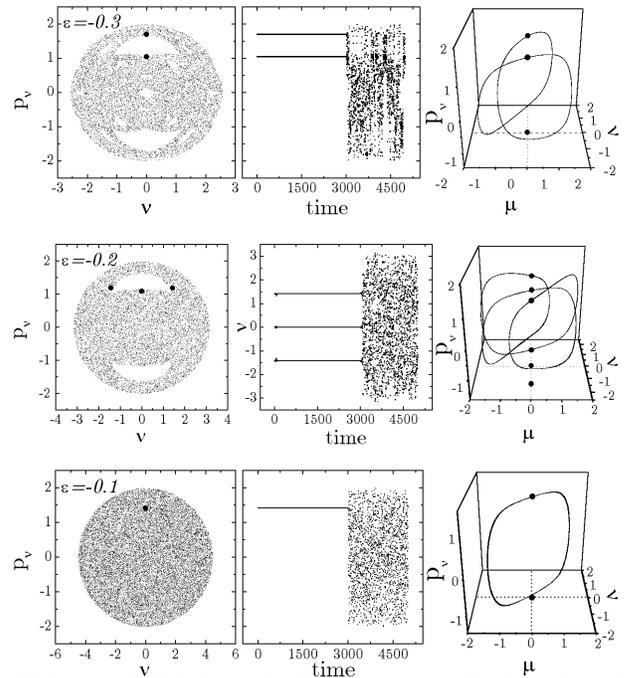}}
\caption{ OGY Control of Diamagnetic Kepler orbits for 3 scaled energies $\epsilon_0 = -0.3, -0.2, -0.1$
and periods $2, 3, 1$ in top, middle ,and bottom pannels:
({\em left}) Poincar\'e section (PS) $\mu= 0, \dot{\mu} > 0$ showing one chaotic trajectory
(filled space) and the controlled UPO (black dots);
({\em middle}) the stabilized $p_{\nu}$ or $\nu$ variable for the first 
3000 intersections with the PS before control is turned off;
({\em right}) corresponding 3D stabilized trajectory. }
\label{dkp_p2_03}
\end{figure}

\begin{figure}[h]
\epsfxsize=.45\textwidth\leavevmode
\centerline{\epsffile{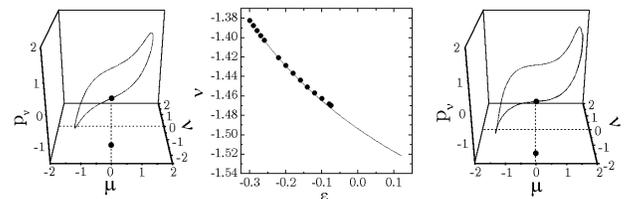}}
\caption{({\em middle}) Tracking of an unstable fixed point for - 0.3$\leq \epsilon \leq $- 0.075. Solid line: AT. ${\Large \bullet}$: STA.  
({\em left}) 3D trajectory for $\epsilon = -0.3$. 
({\em right}) 3D trajectory for $\epsilon = -0.075$. The initial condition for this trajectory is 
that obtained by adiabatic tracking, i.e. ($\nu ^* = -1.46942$ , $p_{\nu}^* = -0.000133596$).}
\label{track}
\end{figure}

\end{document}